# Utilization of e-Nose Sensory Modality as Add-On Feature for Advanced Driver Assistance System

**Ira C. Valenzuela[1], Lean Karlo S. Tolentino[2], Ronnie O. Serfa Juan[3]**

[1]Technological University of the Philippines, Philippines, ira_valenzuela@tup.edu.ph

[2]Technological University of the Philippines, Philippines, leankarlo_tolentino@tup.edu.ph

[3]Cheongju University, South Korea, ronnieserfajuan@cju.ac.kr

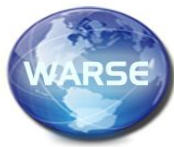

## ABSTRACT

The frequent usage of sensory modalities for Advanced Driver Assistance System (ADAS) and in some In-Vehicles Information System (IVIS) are only for visuals and hearing applications. Yet, the air quality inside the car is not usually monitored. Not to mention, the existing breath analyzers are used only by road traffic police enforcer randomly in examining the alcohol content in motorist's breath to avoid drunk driving related accidents. In this study, we proposed the development and utilization of an e-nose sensory module as an added feature for Controller Area Network (CAN) technology application. An electronic nose is characterized by its capability to intelligently sensed the gases present in the surrounding using an array of gas sensors with a pattern reorganization component. This e-nose comprises of a sensor array of five commercially available sensors (TG 2602, TG 822, TG 825, TG 813 and TG 880), a data acquisition interface, and a microprocessor. A software from the CAN module verified the system functionality. Experimental results indicate that the proposed system is able to identify the five different types of gases namely methane, ethanol, propone, isobutane, and hydrogen with high efficiency. Thus, it can be used as an added safety feature for vehicles.

## 1. INTRODUCTION

Development in microelectronics and embedded system technology have accelerated the change of Driver Assistance Systems towards in driving safety and comfort. The technological and progress, however, results into a more complex, intricate systems and data overload. Assessing designs for automotive services is an emerging and vital field of research.

Today's most frequent usage of sensory modalities for Advanced Driver Assistance System (ADAS) and in some In-Vehicle Information System (IVIS) are only for visuals and hearing applications [1],[2]. These systems have been developed for providing support for driver, but their operation is only concentrated and focused on human/driver's environment only: monitoring of light reflection, night vision scheme, and changing light condition for visual channel, or passenger communication, traffic noise, and mobile calls for auditory notification channel. Implementation for human's secretion like breath analyzer system that monitors driver's breathe composition which contains alcohol on certain level that leads to vehicular accidents is not yet enforced in some ADAS or even in IVIS. Not to mention, the existing breath analyzers are used only by road traffic police enforcer randomly to drivers in examining motorist's breath for alcohol content to avoid drunk driving accidents. Also, inside the car, the air quality is not usually being monitored. one of a common cause of death of drivers is the carbon monoxide poisoning. The causes of this are leaking carbon monoxide from the engine bay, defective exhaust system and others [3].

The main purpose of this study is the development of an electronic nose which will be able to monitor the presence of the five (5) gases namely methane, ethanol, propone, isobutane, and hydrogen inside the car. This will be used as an add-on feature in smart vehicles to ensure safety and to save human lives.

The paper is organized as follows: Section 2 presents the principle behind the e-nose technology and some existing related safety features of ADAS including some discussion of related works regarding e-nose applications which can be adapted for ADAS safety features. Section 3 discusses the proposed system. Then in Section 4, simulations, testing and evaluation results are presented. Finally, Section 5 concludes this paper.

## 2. PRINCIPLE OF OPERATION OF E-NOSE AND SOME EXISTING ADAS SAFETY FEATURES

### 2.1 E-nose and its Principle of Operations

An array of sensors can imitate the mammalian capability to smell to respond to different aromas and this is called electronic nose [4]. There are different techniques in drawing the odor molecules namely: diffusion method and pre-concentrators [5]. Conductivity is equivalent to the physical and chemical changes in the array of sensors while sensing the induced odor [6].

Nowadays, an e-nose is composed of an array of gas sensors and a microprocessor unit that has a classification or recognition algorithm and lookup table that is embedded on it. The array of sensors typically reacts to a variety of chemicals and selectively identifying the mixture of odor samples [7].

Each "cell" in the array can behave like a receptor which responds to different odors to varying degrees [8]. Figure 1 shows the sensing method of an e-nose. The odor molecules serve as the input of the system in which it will be processed through converting the molecules to electrical signals. These signals are then recognized by the pattern recognition system. This is designed with uniqueness in identifying the odor to a group of odors.

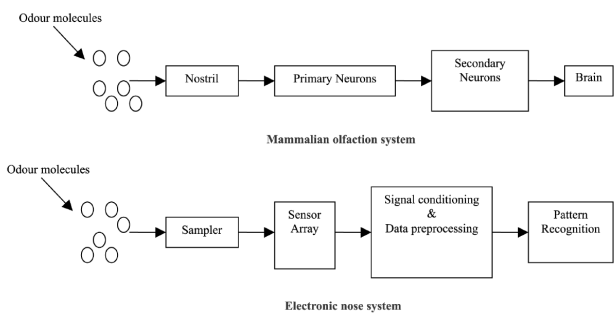

**Figure 1:** Electronic Nose System

Gas sensor are usually made of the following materials: metal-oxide semiconductor filed effect transistor (MOSFET), conducting polymers, surface acoustic waves (SAW), polymer composites, electrochemical gas sensors, quarts crystal microbalance, fiber optic sensors and others [9]. Commercially available sensors are commonly used as part of the sensor array of an e-nose such as Micronas which is based on capacitive effect and other gas sensor which is based on resistive effect [10].

## 2.2 E-nose Sensor Response to Odorants

In general, a first order time response can be observed in the sensing capability of an e-nose [4]. In odor analysis, a baseline has to be developed first by flushing a reference gas in the sensor. This causes a change in the signal until it reaches the steady state. Then it will be flushed out and will return to its baseline as shown in Figure 2. Response time is defined as the time duration in which the sensor is exposed to odor. On the other hand, the recovery time is defined as the time duration when the sensor returns to its baseline [4]. Finally, the response of the sensor to odor is analyzed.

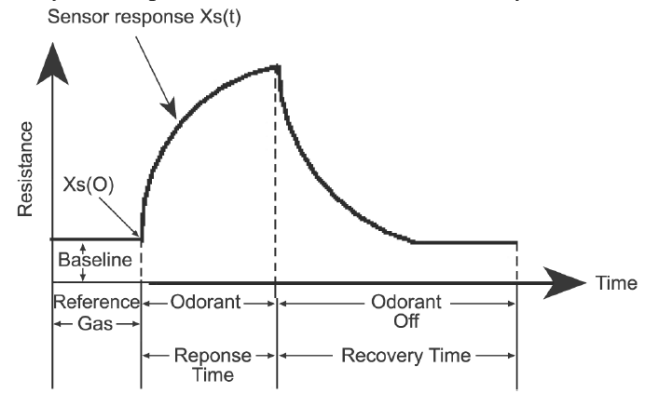

**Figure 2:** E-Nose Response to Odorant

## 2.3 Related Works

There are only limited proposed papers that attempts to utilize the e-nose system in the ADAS technology. An example of this is the car cabin measurement of the air quality. It is important to be controlled as this way will alert the driver regarding the danger of poor air quality [11]. Also, the different gas sensors are compared for the possible use as an add-on feature in ADAS and verify its effectivity. Another study conducted the application of intelligent alcohol detection in a car using MQ-2 gas sensor [12]. This gas sensor analyzes the gas content in the car. Consequently, the sensor output voltage will increase as an input for the actuator. However, the use of single gas sensor might be disadvantageous due to limited specification.

Another application of an e-nose system is in automotive fuel qualification through classifying the different types of gasoline, diesel, and heating oil. The system is able to provide the exceeding values for the volatile organic compounds (VOCs) found in fuels [13]. However, this system requires the stabilization of a number of essential measurement parameters. Moreover, supervised learning algorithm such as ANN has been used as gas detection system for vehicle exhaust. The drawback of this present system is the delay between successive tests [14].

With the advanced sensing module, sophisticated but useful algorithms, and increasing computational resources, this proposed system aims to leverage the usage of e-nose system which enhance the advanced driver assistance systems (ADAS).

## 3. PROPOSED SYSTEM

Figure 2 shows the proposed e-nose chamber system. The chamber is composed of various gas sensors in modular format for different types of gas inputs. The conductivity of the sensor increases as the gas concentration in the chamber increases.

This conductivity is equivalent to the gas concentration.

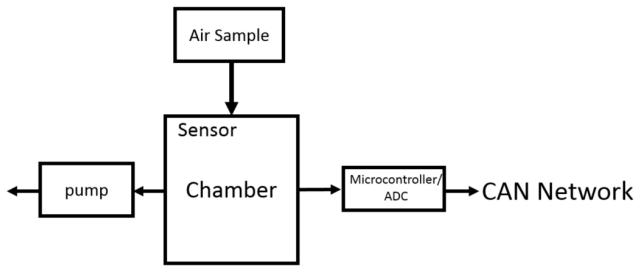

**Figure 3:** Proposed E-Nose System Block Diagram

Figure 4 shows the gas detector circuit using the Figaro gas detector sensors. This sensor is made up of tin dioxide ($SnO_2$) semiconductor material. It is characterized by its low conductivity in clean air. When it comes in contact with any gaseous elements, its internal resistance immediately drops [9]. This sensor is coupled to the proposed e-nose chamber.

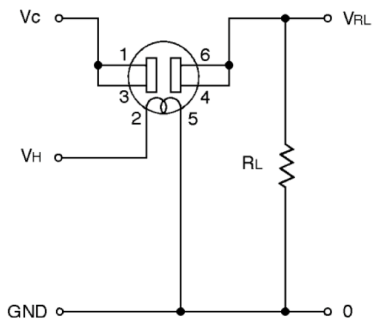

**Figure 4:** Schematic Diagram of Figaro Gas Sensor

Figure 5 shows the actual image of the Figaro TGS 813 gas sensor. This sensing element is a general-purpose sensor which has a noticeable sensing characteristic to a variety of gases. Also, its specification was designed to operate with a stabilized 5V heater supply. Normally, this sensor is used in the detection of methane, propane and butane which is normally utilized in domestic applications.

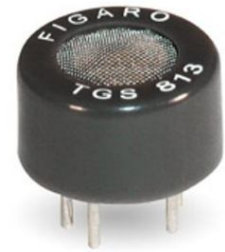

**Figure 5:** Actual Image of Figaro TGS 813

The response of sensor TGS 813 to gases in the surrounding is shown in Figure 6.

It can be seen that this sensor stabilizes in a short period of time which is appropriate and best as a sensing device for e-nose. This can gases present in the environment during measurements.

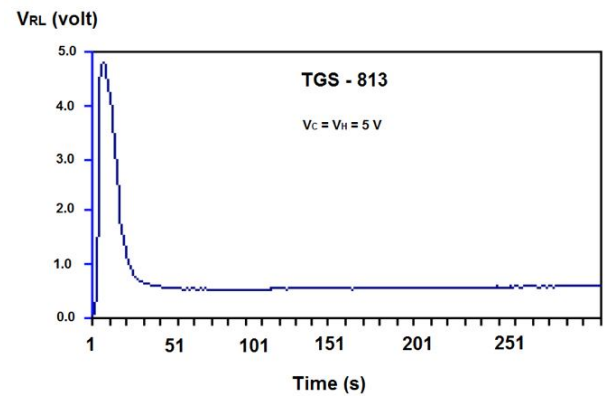

**Figure 6:** Figaro TGS 813 Sensor Response Time

The sensing unit in the chamber consists of a temperature sensor, five (5) Metal Oxide Gas sensors, and a humidity sensor that serves as an electrical interface for conditioning the sensor measurement. In obtaining the chemical operating point, the sensor must be heated non stop at 300 °C. Figure 4 illustrates the circuit used in measuring the conductivity of the sensors as it relates to the concentration of the gases. The circuit contains the load resistance ($R_L$), bias voltage of the sensing element ($V_C$), bias voltage of the heating element ($V_H$) and the output voltage ($V_{RL}$).

Furthermore, each of the sensors is aimed to measure particular gases. However, they have some responses to other gases which a small limitation on their characteristic. Some MOSFET gas sensors are joined to form an array and improves the ability to detect odors because odor is a combination of different gases. The MOSFET gas sensors included in this proposed system are listed in Table 1.

**Table 1:** MOSFET Gas Sensors

| Types of Figaro Sensor | Types of Gases |
|---|---|
| 2602 | Contaminants in the air |
| 822 | Organic Solvent Vapor |
| 825 | H$_2$S |
| 813 | Combustible Gas |
| 880 | Oil Vapor |
| SHT 71 | Temperature and Humidity |

Figure 7 shows the overall system of the proposed e-nose module.

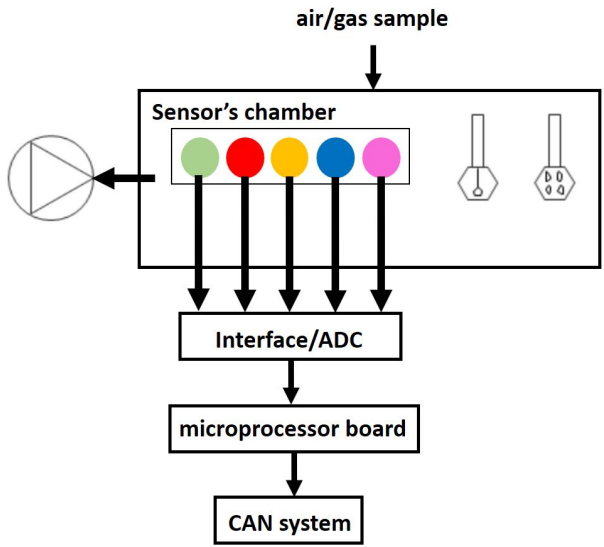

**Figure 7:** Overall System of the Proposed Module

## 4. RESULTS AND DISCUSSION

To gain the reliability of the proposed system, an experimental testing was conducted. The setup components of the proposed system comprise of a sensor array, interface circuit, a microprocessor board, sample of gases, a gas pump, a verification monitor which is connected in the CAN network. This was tested in a secure and controlled environment with proper ventilation and air filter to avoid hazardous effect to the researchers. It was tested with five types of gases, namely, methane, ethanol, propone, isobutane, and hydrogen.

The testing procedures were as follows:
(1)  The training and testing module is composed of a raspberry pi capable PiCAN 2 board provides CAN-Bus with MCP2515 CAN controller and MCP2551 CAN transceiver.
(2) Valve 1 is connected at the opening of the chamber and valve 2 is placed between the chamber and gas pump.
(3) Valve 1 is shut, and Valve 2 is released. The vacuum pump is set to pump the gas in 10 seconds time range.
(4) Valve 1 is opened to connect in the chamber opening for

the gas sample. Valve 2 is shut while the vacuum pump is turned off for 10 seconds.
(5) Valve 1 is closed, and the sensors resistance has allotted a 1-minute time range to obtain a steady state. The sensors characteristic values are displayed in the CAN system.
(6) The sensor chamber is disconnected from the gas sample. Valve 1 is replenished by a clean air. Valve 2 is released, the odor has been discharged, and the system is aired out with clean air for a certain time before testing the other gases.

The gas inflow is regulated using mass flow controllers (MFCs) with the allowable maximum value. The mass flow in the path uses the standard cubic centimeters per minute (sccm) as the flow measurement. Table 2 shows the maximum allowed value of the MFC.

**Table 2:** MFC Usage and its Maximum Values

| Gas/Contaminant | Maximum Value (sccm) |
|---|---|
| CH4 | 200 |
| C2H5OH | 200 |
| C3H8 | 200 |
| C4H10 | 200 |
| H | 200 |

It can be seen in Figure 8 the TGS 813 relative sensitivity to different types of gases in ideal specification. The graph shows the relationship of the gas concentrations to the sensor resistance. From this, it can be observed that as the concentration of the gases increases, the Rs/Ro decreases. Thus, giving the implication of inversely proportional relationship.

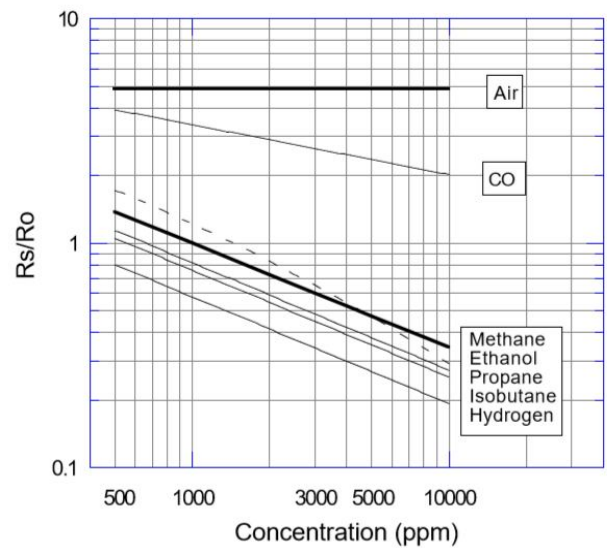

**Figure 8:** Sensitivity to Various Gases (Rs/Ro)

Likewise, the actual testing was conducted to compare the results to the ideal characteristics of the sensing elements. Figure 9 shows the results of the actual testing. It can be seen that the proposed E-nose has the same characteristics as its ideal response.

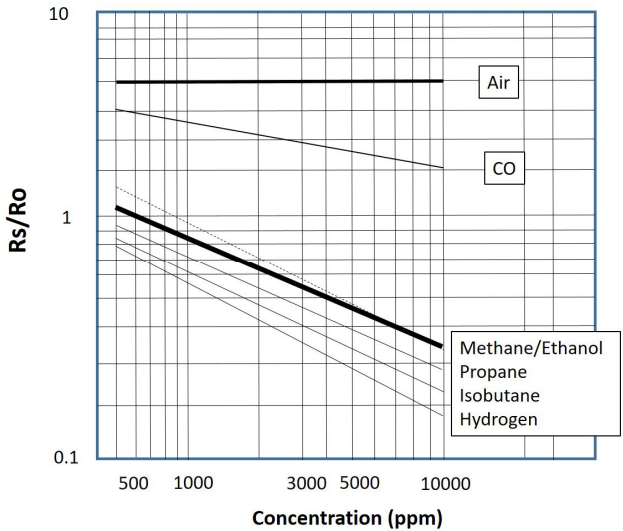

**Figure 9:** Actual Sensitivity to Various Gases (Rs/Ro)

Figure 10 served as the reference or baseline for characterizing the actual response of the developed system. And Figure 11 shows the actual sensitivity characteristics to the different gases. The plot suggests that it has the same characteristics as its ideal response.

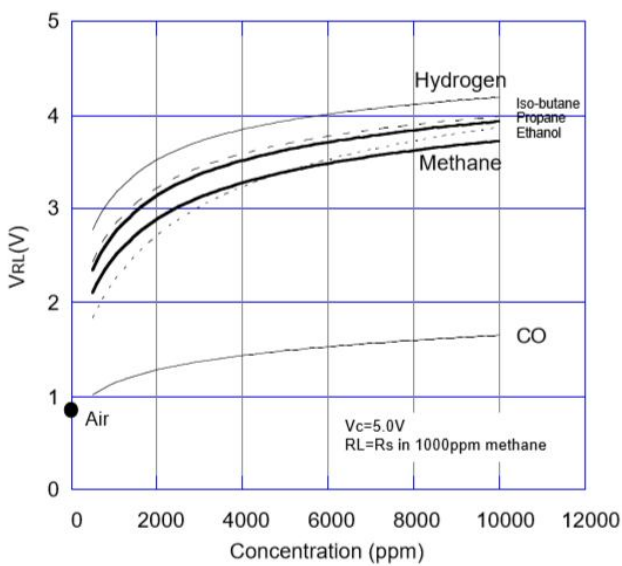

**Figure 10:** Sensitivity to Various Gases ($V_{RL}$)

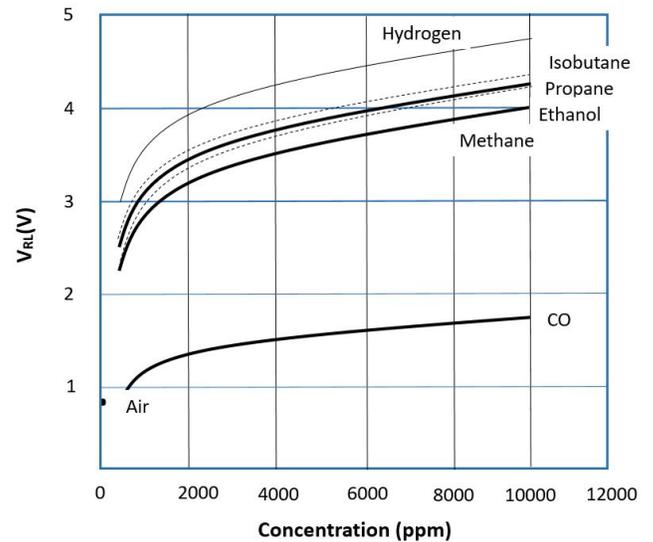

**Figure 11:** Actual Sensitivity to Various Gases ($V_{RL}$)

## 5. CONCLUSION

A chemical sensor array system like an e-nose module may detect and identify any samples through provisions of certain criteria. This array should be sensitive enough to the samples of interest and selective enough to be able to distinguish between them and all others. The differential responsiveness of e-nose to various gases had given a noticeable advantage over the traditional analytical instruments. Quick results, high sensitivity, non destructive type of sampling and no chemicals needed are few of the advantages of this system.

Based on the experiments done and the data gathered, it can be concluded that the proposed system was able to detect the different gases namely methane, ethanol, propone, isobutane, and hydrogen present inside the car. The response of the device clearly demonstrates that this can be used as an add-on feature for advanced driver assistance system.

Not all technologies provide positive effect to human and to the environment. However, in future, the improvements in e-nose applications are promising. These existing drawbacks can lead researchers in many fields become more aware of its capabilities and limitations. This proposed paper, aims to optimize on what advantages can an electronic nose provides to ADAS technology. With the advent of the applications of Fourth Industrial Revolution (FIRe), new potential development in e-nose sensors will help to realize innovative products and advanced systems and will lead to the solutions in solving new problems which will be beneficial to mankind.

## ACKNOWLEDGEMENT

The authors would like to thank the Technological University of the Philippines for funding this research and the Cheongju University for allowing the researchers to use their equipment during testing.